\documentclass[aps,prl,twocolumn,groupedaddress,showpacs,notitlepage]{revtex4}
\usepackage{graphicx}
\usepackage{amsmath,amssymb}
\usepackage{color}
\usepackage[dvipdfmx]{hyperref}

\newcommand{\bra}[1]{\left<#1\right|}
\newcommand{\ket}[1]{\left|#1\right>}
\newcommand{\pd}{{\phantom{\dag}}}
\newcommand{\abs}[1]{\left|#1\right|}

\begin{document}
\title{Resonant Inelastic X-ray Scattering on Spin-Orbit Coupled 
Insulating Iridates}
\author{Luuk J.P. Ament$^1$, Giniyat Khaliullin$^2$, Jeroen van den Brink$^3$}
\affiliation{$^1$Institute-Lorentz for Theoretical Physics, Universiteit Leiden, 2300 RA Leiden, The Netherlands}
\affiliation{$^2$Max-Planck-Institut f\"ur Festk\"orperforschung, Heisenbergstrasse 1, D-70569 Stuttgart, Germany}
\affiliation{$^3$Leibniz-Institute for Solid State and Materials Research Dresden, D-01171 Dresden, Germany}
\date{\today}

\begin{abstract}
We determine how the elementary excitations of iridium-oxide
materials, which are dominated by a strong relativistic spin-orbit coupling, appear in Resonant Inelastic X-ray Scattering (RIXS).  Whereas the RIXS spectral weight at the $L_2$ x-ray edge vanishes, we find it to be strong at the $L_3$-edge. Applying this to Sr$_2$IrO$_4$, we observe that RIXS, besides being sensitive to local doublet to quartet transitions, meticulously maps out the strongly dispersive delocalized excitations of the low-lying spin-orbit doublets. 
\end{abstract}

\pacs{
78.70.Ck, 
75.30.Et, 
75.25.Dk, 
71.70.Ej 
}

\maketitle

Materials containing ions with an orbital degeneracy show a plethora
of physical effects related to the spontaneous lifting of this
degeneracy~\cite{Tokura2000}. The two degeneracy-breaking mechanisms
in Mott insulators traditionally considered are the cooperative
Jahn-Teller effect~\cite{Jahn1937}, involving a change in lattice
symmetry, and the superexchange interactions~\cite{Kugel1982},
intertwining long-range ordering of orbital and magnetic degrees of
freedom. A third and less explored possibility exists in systems
containing heavy ions, where strong relativistic spin-orbit coupling
dominates the spin-orbital physics~\cite{Kim2009}.

The strong spin-orbit interaction can cause entirely new kinds of
ordering that are of  topological nature. This was recently proposed for certain
iridium-oxides~\cite{Shitade2009,Pesin2010}, members of a large family
of iridium-based materials. 
Na$_2$IrO$_3$, for instance, is predicted to be a
topological insulator exhibiting the quantum spin Hall effect at room
temperature~\cite{Shitade2009}. The topologically non-trivial state
arises from the presence of complex hopping integrals,  resulting from
the unquenched iridium orbital moment. This system can also be
described in terms of a Mott insulator, with interactions between the
effective iridium spin-orbital degrees of freedom goverened by
the Kitaev-Heisenberg model~\cite{Jackeli2009,Kitaev2006,Chaloupka2010}. In the
pyrochlore iridates A$_2$Ir$_2$O$_7$ (where A is a $3+$ ion), a quantum phase transition from a topological band insulator to a topological Mott insulator has been proposed as a function of the electron-electron interaction strength~\cite{Pesin2010,Yang2010,Wan2010}.

To establish whether and how such novel phases are realized in iridium
oxides it is essential to probe and understand their spin-orbital
ordering and related elementary excitations. In this context it is
advantageous to consider the structurally less complicated,
single-layer iridium perovskite Sr$_2$IrO$_4$. This material is in
many respects the analog of the high-$T_c$ cuprate parent compound
La$_2$CuO$_4$~\cite{Jackeli2009}. Structurally it is identical,
with the obvious difference that the Ir $5d$ valence electrons are, as
opposed to the Cu $3d$ electrons, very strongly spin-orbit coupled.
The similarity cuts deeper, however, as the low energy sector of the
iridates is spanned by local spin-orbit doublets with an effective
spin of 1/2, which reside on a square lattice and interact via
superexchange -- a close analogy with the undoped cuprates. This
observation motivates doping studies of Sr$_2$IrO$_4$ searching for
superconductivity~\cite{Cosio-Castaneda2007,Klein2008}. Experimentally,
however, far less is known about the microscopic ordering and
excitations in iridates than in cuprates. Inelastic neutron
scattering, which can in principle reveal such properties, is not
possible because Ir is a strong neutron absorber~\cite{Powell1993}
and, moreover, crystals presently available tend to be tiny. As a consequence not even the interaction strength between the effective spins in the simplest iridium-oxides is established: estimates for Sr$_2$IrO$_4$, for instance, range from $\sim$50 meV~\cite{Jackeli2009} to $\sim$110 meV~\cite{Fujiyama}.

In this Letter we show that while for iridates neutron scattering falls short,
photon-in photon-out scattering in the form of resonant inelastic
x-ray scattering (RIXS)~\cite{Kotani2001} fills the void: RIXS at the
iridium $L$-edge offers direct access to the excitation spectrum across
the Brillouin zone, enabling one to measure the dispersion of
elementary magnetic excitations. Besides the low energy magnons
related to long-range order of the doublets, RIXS will also reveal the
dynamics of higher energy, doublet-to-quartet, spin-orbit
excitations. This allows to directly test theoretical models for the
excitation spectra and extract accurate values of the superexchange and spin-orbit coupling constants $J$ and $\lambda$, respectively.

{\it Ir$^{4+}$ ionic ground state}. --- In the iridium-oxides
Ir$^{4+}$ ions are located in octahedra of oxygen ions, splitting the
$5d$ levels by $\sim3$ eV into $e_g$ and $t_{2g}$
orbitals~\cite{Moon2006}. Because this crystal field splitting is an
order of magnitude larger than the spin-orbit coupling $\lambda$, the
$t_{2g}$ levels do not hybridize much with the $e_g$ orbitals, and a
$t_{2g}^5$ configuration is established~\cite{Kim2008}. The symmetry
of the $t_{2g}^5$ ground state is in principle governed by three
factors: superexchange interactions, additional lattice-induced
crystal field splittings, and relativistic spin-orbit
coupling~\cite{Khaliullin2005}. The superexchange
$J$~\cite{Jackeli2009} is estimated to be about an order of magnitude
smaller than $\lambda \approx 0.4$ eV~\cite{Schirmer1984}. An elongation of the octahedra along the $z$ axis~\cite{AxesFootnote} favors a ground state where the hole is in the $xy$ orbital. But experimental data strongly favor the spin-orbit coupling scenario over the lattice splitting scenario, however~\cite{Kim2009,Kim2008}.

The orbital degree of freedom of the hole can be described by an
effective angular momentum $l = 1$, related to the true orbital
angular momentum by ${\bf l} = -{\bf L}$~\cite{Kanamori1957}. The
orbital eigenstates of $l_z$ are described by the annihilation
operators $d_{0,\pm 1}$, defined in terms of the real $t_{2g}$ wave functions by the relations
$d_{yz} = -  ( d_1 - d_{-1} )/\sqrt{2}$, $d_{zx} = i ( d_1 + d_{-1})/\sqrt{2}$, $d_{xy} = d_0$. When the spin-orbit coupling term is projected to the $t_{2g}$ subspace, it becomes $-\lambda {\bf l}\cdot {\bf S}$. Tetragonal lattice distortions can also be included, and the Hamiltonian for a single Ir ion is~\cite{Jackeli2009} $ H = -\lambda {\bf l}\cdot {\bf S} - \Delta l_z^2$,  with $\Delta > 0$ for elongation along the $z$ axis. The six eigenstates group into three Kramers doublets, described by the fermions $f$, $g$ and $h$ with annihilation operators
\begin{align}
  &\begin{array}{l} f_\uparrow = \sin \theta\ d_{0\uparrow} - \cos \theta\ d_{1\downarrow}, \\
  f_\downarrow = \cos \theta\ d_{-1\uparrow} - \sin \theta\ d_{0\downarrow}, \end{array} \;\;\;
  \begin{array}{l} g_\uparrow = d_{1\uparrow}, \\
    g_\downarrow = d_{-1\downarrow}, \end{array} \nonumber \\
  &\begin{array}{l} h_\uparrow = \cos \theta\ d_{0\uparrow} + \sin \theta\ d_{1\downarrow}, \\
  h_\downarrow = \cos \theta\ d_{0\downarrow} + \sin \theta\
  d_{-1\uparrow}, \end{array} \label{eq:6_fgh}
\end{align}
and with energies $\omega_f = \lambda/ (\sqrt{2}\tan \theta )$, $\omega_g = -\Delta - \lambda/2$ and $\omega_h = -(\lambda \tan
\theta)/\sqrt{2}$, where $\tan 2\theta = 2\sqrt{2}
\lambda/(\lambda-2\Delta)$. For $\Delta=0$, which corresponds to the cubic, isotropic situation $\sin \theta = \sqrt{1/3}$ and $\cos \theta = \sqrt{2/3}$.
For $\lambda /\Delta \ll 1$, the hole's ground state doublet is $\{\ket{xy \uparrow},\ket{xy \downarrow}\}$. For $\lambda/\Delta \gg 1$, the eigenstates are characterized by the total effective angular momentum ${\bf J}_{\rm eff} = {\bf l} + {\bf S}$. In the ground state, the hole occupies the $f$ doublet ($J_{\rm eff} = 1/2$), which is separated by an energy of $3\lambda/2$ from the $J_{\rm eff} = 3/2$ quartet, which splits into the $g$ and $h$ doublets.

The remaining, two-fold ground state degeneracy cannot be removed by Jahn-Teller distortions because the two states in the ground state Kramers doublet have exactly the same charge
distribution. Superexchange coupling, however, is present in all
iridates, and couples the local doublets, thus dictating the low energy collective dynamics of the material.

{\it RIXS cross section}. --- Resonant Inelastic X-ray Scattering is
particularly suited to probe higher energy magnetic excitations and
dispersions, as demonstrated in the
cuprates~\cite{Ament2009,Braicovich2010a,Braicovich2010b,Guarise2010}. The recent
increase in brilliance of the new generation synchrotron X-ray sources
allows for spectra with tremendous resolving powers, enough to resolve
dispersion up to a few tens of meV. In
RIXS~\cite{Kotani2001,Ament2010}, a photon with momentum ${
  \hbar} {\bf k}$, energy $\omega_{\bf k}$ and polarization ${\boldsymbol
  \epsilon}$ is scattered to ${\hbar} {\bf k}'$,
$\omega_{{\bf k}'}$ and ${\boldsymbol \epsilon}'$, losing momentum
$\hbar {\bf q} = \hbar {\bf k} - \hbar {\bf k}'$ and energy $\omega = \omega_{\bf k} - \omega_{{\bf k}'}$ to the sample. $\omega_{\bf k}$ is tuned to a certain atomic resonance of the material under study, greatly enhancing the scattering cross section. In our case, that will be the Ir $L$-edge: the $2p$ core electron is excited into the empty $5d$ $t_{2g}$ state. After a very short time, another electron from the $t_{2g}$ levels can fall back to the core hole under the emission of an outgoing X-ray. The system is left in an excited state, whose energy and momentum are taken from the scattered X-ray photon, which is measured.

The RIXS cross section is described by the Kramers-Heisenberg
equation~\cite{Sakurai1967}, where the photon absorption and
subsequent emission are governed by the dipole operator ${\cal D} =
\sum_{\bf R} e^{i {\bf k}\cdot {\bf R}}\ \! {\bf
  r} \cdot {\boldsymbol \epsilon}$ acting on all electrons of an Ir$^{4+}$ ion at site ${\bf
  R}$. The phonon polarization is ${\boldsymbol \epsilon}$.

The intermediate state has a filled shell (5d $t_{2g}^6$), so the
dominant multiplet effect comes from the core orbital's spin-orbit
coupling $\Lambda$: the 2p core states split into $J=1/2$ (the $L_2$
edge) and $J=3/2$ states (the $L_3$ edge). Since the $L_2$ and the
$L_3$ edge are separated by $1.6$ keV~\cite{Kim2009}, their
interference is negligible, given the much smaller lifetime broadening
of a few eV~\cite{Krause1979}. Because the 2p core states have the
same angular momenta as the 5d $t_{2g}$ states, we can describe them
with the three fermions $F$, $G$ and $H$, analogous to
Eq.~(\ref{eq:6_fgh}), where we replace $(d_{yz},d_{zx},d_{xy})$ by
$(p_x,p_y,p_z)$ and the parameters $\lambda$, $\Delta$ and $\theta$ by
$\Lambda$, $\delta$ and $\Theta$. The tetragonal distortion $\delta$ is expected to be very small for the deep $2p$ core states.

The lifetime broadening at the Ir L edge is still quite large compared
to the dynamics of the 5d
electrons~\cite{Krause1979,Kim2008}. Therefore, we make the fast
collision approximation $E_i+\hbar \omega_{\bf k} - E_n + i\Gamma
\approx i \Gamma$~\cite{Schuelke2007}. The sum over $n$ in the
Kramers-Heisenberg equation can be performed, and comprises the core states of either the $L_2$ or the $L_3$ edge.
In second quantization, the dipole operators are ${\bf r}\cdot {\boldsymbol \epsilon} = \sum_{\alpha,\beta,\sigma} 
  \bra{5d_\alpha} {\bf r} \ket{2p_\beta} \cdot {\boldsymbol \epsilon}\
  d^{\dag}_{\alpha\sigma}\ p^\pd_{\beta\sigma} + \text{h.c.}$ which can be
  denoted as $(D_2 + D_3) + {\rm h.c.}$, where $D_{2,3}$ are the local
  dipole transition operators for the
$L_2$ and $L_3$ edge, respectively. The RIXS amplitude becomes $A_{\bf
  q} \propto \bra{f} \sum_{\bf R} e^{i{\bf q}\cdot {\bf R}} [D^\dag
({\boldsymbol \epsilon}'^*) D ({\boldsymbol \epsilon})]_{\bf R}
\ket{0}$, where ${\bf R}$ runs over all Ir sites and the RIXS intensity
$I_{\bf q} (\omega) = \sum_f |A_{\bf q}|^2 \delta (\omega - E_f)$.
We rewrite the inelastic scattering operator as
\begin{align}
  D^\dag D &=  \sum_{\sigma \in \{\uparrow,\downarrow\}}   \left[  B^{ff}_{\sigma \sigma}  f^\dag_\sigma f^\pd_\sigma + B^{ff}_{\sigma \bar{\sigma}}  f^\dag_\sigma f^\pd_{\bar{\sigma}} + B^{fg}_{\sigma \sigma} f^\dag_\sigma g^\pd_\sigma  \right.   \nonumber \\
    & \ \ \ \ \ \ \ \ \ \ \  \   \left.   + B^{fg}_{\sigma \bar{\sigma}} f^\dag_\sigma
      g^\pd_{\bar{\sigma}}+ B^{fh}_{\sigma \sigma}  f^\dag_\sigma h^\pd_\sigma + B^{fh}_{\sigma \bar{\sigma}}  f^\dag_\sigma h^\pd_{\bar \sigma} \right]. \label{eq:O2}
\end{align}
Integrating out the core hole degree of freedom one obtains at the $L_2$ edge the intra-doublet scattering matrix elements
$
B^{ff}_{\sigma \sigma} =-\sin^2 (\theta - \Theta)  \epsilon'^*_{\bar{\sigma}} \epsilon^\pd_{\bar{\sigma}}
$
and 
$
B^{ff}_{\sigma \bar{\sigma}} =0. 
$
The doublet-quartet excitation matrix elements of the spin-orbit multiplet are
$
B^{fg}_{\sigma \sigma}= \sin (\theta - \Theta) \cos \Theta\ \epsilon'^*_z \epsilon^\pd_{\bar{\sigma}},
$
$
B^{fg}_{\sigma \bar{\sigma}}=-(-1)^\sigma \sin (\theta - \Theta) \sin \Theta\ \epsilon'^*_\sigma \epsilon^\pd_{\bar{\sigma}},
$
$
B^{fh}_{\sigma \sigma}= -\frac{1}{2} (-1)^\sigma \sin 2(\theta - \Theta) \epsilon'^*_{\bar{\sigma}} \epsilon^\pd_{\bar{\sigma}}$ 
and 
$B^{fh}_{\sigma \bar{\sigma}}=0$,
where $(-1)^\sigma$ is $1$ for $\sigma =\ \uparrow$
and $-1$ for $\sigma =\ \downarrow$. Further, $\epsilon_\uparrow =
\epsilon_+$ and $\epsilon_\downarrow = \epsilon_-$, with $\epsilon_\pm
= (\epsilon_x \pm i\epsilon_y)/\sqrt{2}$. In the case of dominant
spin-orbit coupling, $\theta = \Theta$. Since all matrix elements at
the $L_2$ edge are proportional to $\sin (\theta-\Theta)$, the
inelastic scattering intensity vanishes completely in this case, in
addition to a vanishing of the elastic intensity~\cite{Kim2009}.

At the $L_3$ edge, however, RIXS is fully allowed. The matrix
elements are:
$
B^{ff}_{\sigma \sigma} =- \sin^2 \theta\ \epsilon'^*_\sigma \epsilon^\pd_\sigma
    - \cos^2 (\theta-\Theta) \epsilon'^*_{\bar{\sigma}}
    \epsilon^\pd_{\bar{\sigma}} - \cos^2 \theta\ \epsilon'^*_z \epsilon^\pd_z
$
and 
$
B^{ff}_{\sigma \bar{\sigma}} = \frac{1}{2} (-1)^\sigma \sin 2\theta\ (\epsilon'^*_{\bar{\sigma}}
    \epsilon^\pd_z-\epsilon'^*_z \epsilon^\pd_\sigma)
$
for the intra-doublet ones, and 
$
B^{fg}_{\sigma \sigma}= \cos (\theta-\Theta) \sin \Theta\ \epsilon'^*_z \epsilon^\pd_{\bar{\sigma}},
$
$
B^{fg}_{\sigma \bar{\sigma}}= (-1)^\sigma \cos (\theta-\Theta)  \cos \Theta\ \epsilon'^*_\sigma  \epsilon^\pd_{\bar{\sigma}},
$
$
B^{fh}_{\sigma \sigma}= \frac{1}{2} (-1)^\sigma [
    \sin 2(\theta-\Theta) \epsilon'^*_{\bar{\sigma}} \epsilon^\pd_{\bar{\sigma}}
    - \sin 2\theta\ (\epsilon'^*_\sigma \epsilon^\pd_\sigma
    -\epsilon'^*_z \epsilon^\pd_z)] 
$,
$
     B^{fh}_{\sigma \bar{\sigma}}= -\sin^2 \theta\ \epsilon'^*_z \epsilon^\pd_\sigma 
     -  \cos^2 \theta\ \epsilon'^*_{\bar{\sigma}} \epsilon^\pd_z
$
for the doublet-quartet excitations. For excitations within the $J_{\rm eff} =
1/2$ doublet, the scattering operator can be
rewritten in terms of the effective angular momentum operator, which in
the limit $\Delta/\lambda \ll 1$ takes the particularly simple form $D^{\dag}_3D^\pd_3
= \frac{2}{3} \left( {\boldsymbol \epsilon}'^* \cdot {\boldsymbol
    \epsilon}\ \openone + {\bf P}\cdot {\bf J}_{\rm eff}
\right)$, where $P_x = i (\epsilon'^*_y \epsilon_z - \epsilon'^*_z \epsilon_y)$
and its cyclic permutations $P_{y,z}$ are polarization factors. Here, the first term corresponds to elastic scattering while
the ${\bf P}\cdot {\bf J}_{\rm eff}$ term gives rise to inelastic scattering.

\begin{figure}
      \centering
      \includegraphics[width=\columnwidth]{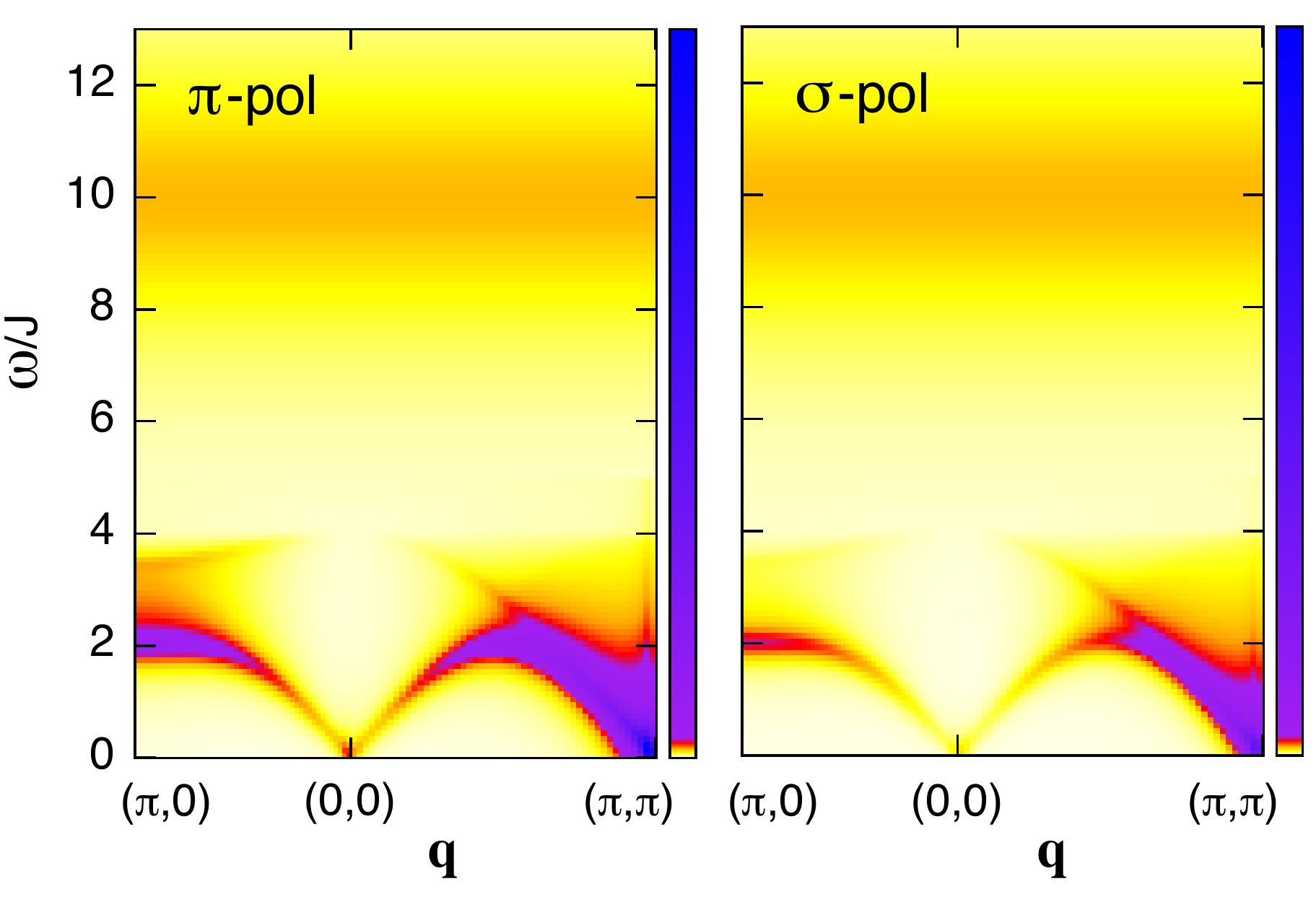}
      \caption{(Color online.) RIXS spectra of Sr$_2$IrO$_4$ at the Ir $L_3$ $t_{2g}$
        edge. The left panel shows the spectrum for incoming $\pi$ polarization, and the
        right one for incoming $\sigma$ polarization. The outgoing
        polarization is not measured. The intra-doublet excitations
        are broadened by $J/10$. \label{fig1}}
\end{figure}

{\it RIXS on Sr$_2$IrO$_4$.} --- Up to this point, the discussion is general and applies to all
materials with an Ir$^{4+}$ ion in an octahedral crystal
field. Calculation of the RIXS spectra for a particular iridate is
straightforward given the Hamilonian that captures the interactions
between the Ir degrees of freedom. In Sr$_2$IrO$_4$ the effective low
energy Hamiltonian is obtained from the spin-orbital superexchange for the triply degenerate $t_{2g}$ orbitals
(Eq.~(3.11) from Ref.~\cite{Khaliullin2005}) by projecting it on the low energy Kramers doublet. In the case of strong spin-orbit coupling, one finds a Heisenberg Hamiltonian for these pseudo-spin-$1/2$ states, with weak dipolar anisotropy due to Hund's rule coupling. The rotation of the octahedra around the $z$ axis over an angle $\alpha \approx 11^\circ$ introduces a Dzyaloshinsky-Moriya interaction, but after an appropriate spin rotation the Hamiltonian remains of Heisenberg type~\cite{Jackeli2009}.

At the Ir $L_3$ edge, excitations within the $J_{\rm eff} = 1/2$
doublet can be described in terms of Holstein-Primakoff bosons. The
single- and double-magnon intensities are, respectively,
\begin{align}
  I^{(1)} &\propto \left[ \abs{\frac{\sin \alpha}{\sqrt{2}} (P_x+P_y) (u_{\bf q}+v_{\bf q}) - iP_z (u_{\bf q}-v_{\bf q})}^2 \right. \nonumber \\
  &\;\;\;\; \left. + \frac{1}{2} \cos^2 \alpha\ \abs{P_x-P_y}^2(u_{\bf q} - v_{\bf q})^2 \right] \delta (\omega - \omega_{\bf q}), \nonumber \\
  I^{(2)} &\propto \frac{2}{N} \sum_{\bf k} \left[ \sin^2 \alpha \abs{P_x-P_y}^2 \left( u_{{\bf k}+{\bf q}} v_{\bf k} + u_{\bf k} v_{{\bf k}+{\bf q}} \right)^2 \right. \nonumber \\
  &\;\;\;\; \left. + \cos^2 \alpha \abs{P_x+P_y}^2 \left( u_{{\bf k}+{\bf q}} v_{\bf k} - u_{\bf k} v_{{\bf k}+{\bf q}} \right)^2 \right] \nonumber \\
  &\;\;\;\; \times \delta (\omega - \omega_{{\bf k}+{\bf q}} - \omega_{\bf k}), \label{eq:I2}
\end{align}
with $u_{\bf k} =  (1/\sqrt{1-\gamma^2_{\bf
    k}} + 1)^{1/2} /{\sqrt{2}}$, $v_{\bf k} = 
(1/\sqrt{1-\gamma^2_{\bf k}} - 1)^{1/2}  {\rm sign}
(\gamma_{\bf k})/{\sqrt{2}}$ and $\gamma_{\bf k} = (\cos k_x + \cos k_y)/2$. 
A remarkable difference with $L$-edge RIXS on cuprates~\cite{Ament2009} is that the large spin canting, reflected in the appreciable value of $\alpha$, causes the presence of spectral weight in the center of the Brillouin-zone, at ${\bf q} = {\bf 0}$.

Transitions from $J_{\rm eff} = 1/2$ to $3/2$, which are at an energy
of $\tfrac{3}{2}\lambda$, are expected to show a less pronounced ${\bf q}$ dependence. The
crystal field splitting of the quartet states is probably too small to
resolve with current RIXS instruments, so we give the integrated
intensity of all these excitations: $I^{(g+h)} \propto 2 +
\abs{{\boldsymbol \epsilon}' \cdot {\boldsymbol \epsilon}}^2 -
\abs{{\boldsymbol \epsilon}'^* \cdot {\boldsymbol
    \epsilon}}^2.$ The polarization terms cancel unless both incoming
and outgoing X-rays are circularly polarized.

\begin{figure}
      \centering
      \includegraphics[width=\columnwidth]{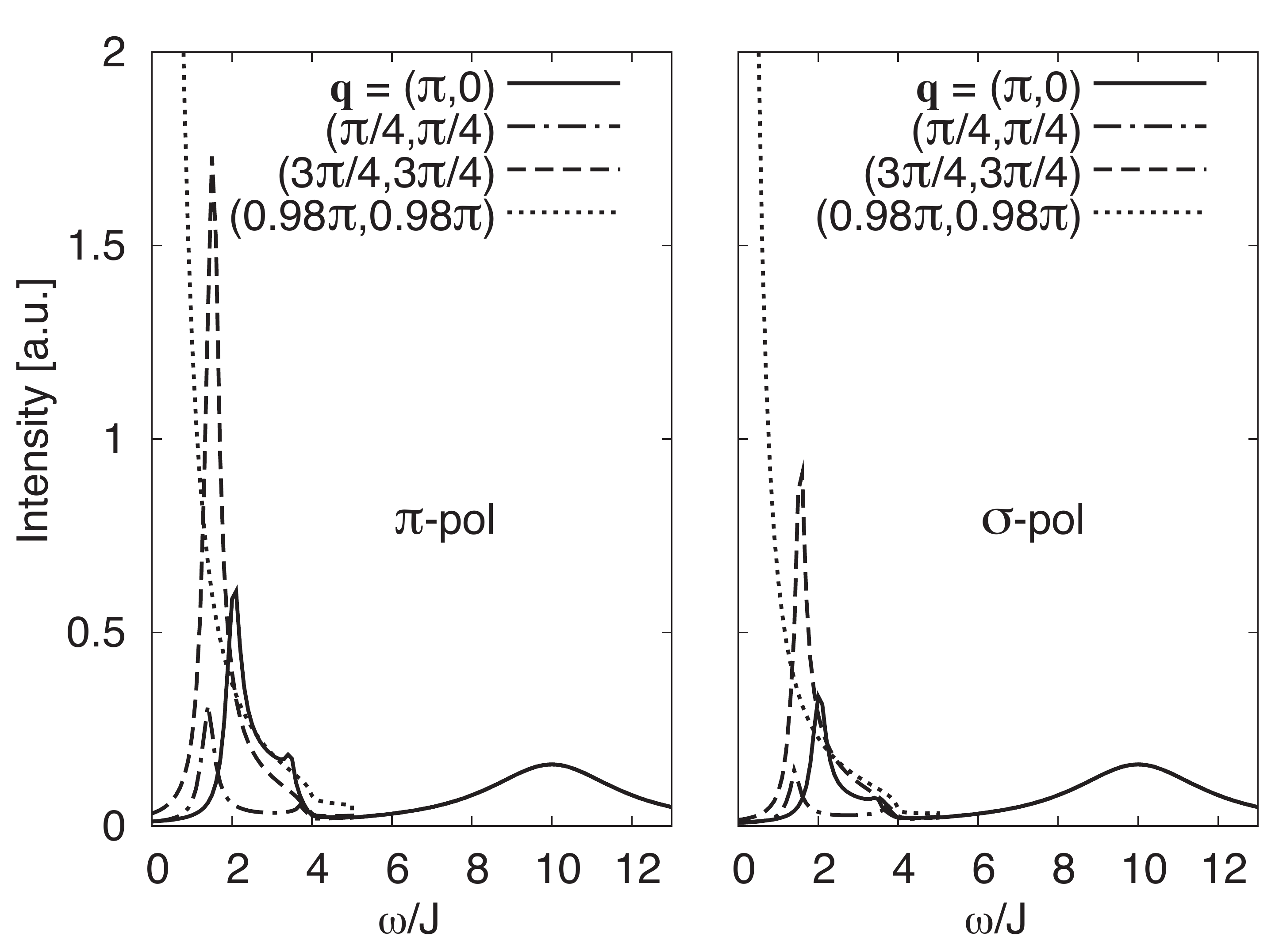}
      \caption{Vertical cuts through Fig.~\ref{fig1}. The left panel
        shows spectra at several ${\bf q}$ for incoming $\pi$ polarization, and the
        right one for incoming $\sigma$ polarization. \label{fig2}}
\end{figure}

{\it Computed RIXS spectra}. --- We now evaluate the different
contributions to the RIXS intensity. Single and double magnon
contributions $I^{(1,2)}$ and those from the $J_{\rm eff} = 1/2$ to $3/2$ excitations $I^{(g+h)}$ are presented for the specific
case of a $90^\circ$ scattering angle with the scattering plane
perpendicular to the IrO$_2$ layers, and ${\bf q}$ in
the first (2D) Brillouin zone. The resulting cross sections are shown
in Figs.~\ref{fig1} and \ref{fig2}, where we used
$\tfrac{3}{2}\lambda/J = 10$. The low-energy intra doublet excitations
with $\Delta J_{\rm eff} = 0$ show a very distinct magnon dispersion,
the intensity of which is strongly varying with $\bf q$. The
doublet-quartet transitions with locally $\Delta J_{\rm eff} = 1$ are
at $\tfrac{3}{2} \lambda$, corresponding to $0.5$ - $0.6$ eV~\cite{Schirmer1984}. This implies
they  match in energy the large spectral weight charge modes
observed in optical absorption in
Sr$_2$IrO$_4$~\cite{Moon2009}. Even if the local multiplet excitations
are not optically active themselves, there will be strong mixing of the
spin-orbit excited state with inter-site charge excitations
across the Mott gap. This causes a delocalization of the doublet-quartet
mode on the scale of the intersite hopping $t$. The dispersion and momentum dependent
spectral weight modulations that this causes is beyond the present model
calculations; here it only reflects in the use of an effective
broadening of the doublet-quartet mode with $t$, corresponding to about $4J$.

\begin{figure} 
      \centering
      \includegraphics[width=.8\columnwidth]{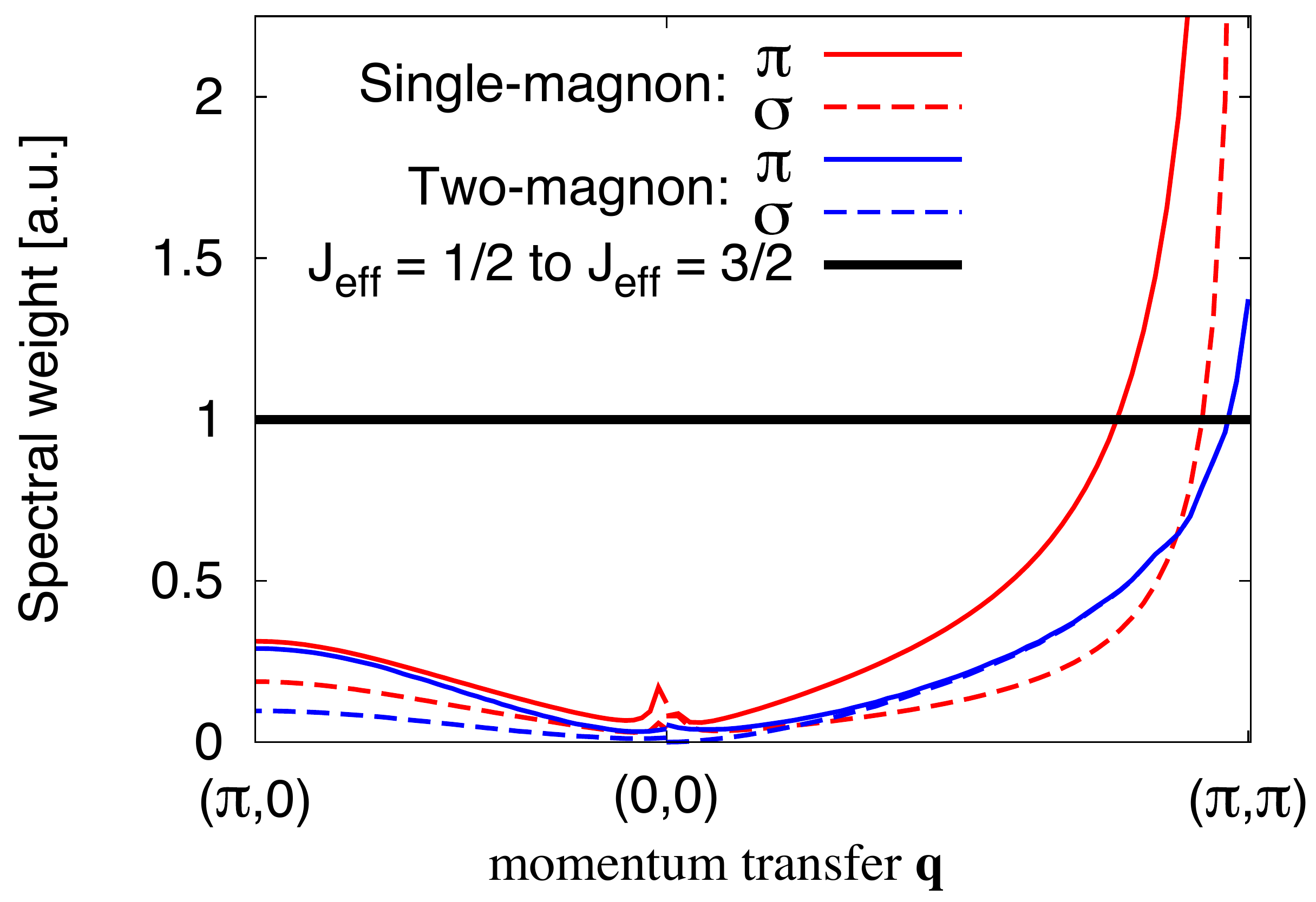}
     \caption{(Color online.) Spectral weight of different excitations. The units on the vertical axis are chosen such that the $J_{\rm eff} = 1/2$ to $3/2$ excitation has a spectral weight of unity. \label{fig3}}
\end{figure}

To summarize, we have determined the effective scattering operators for direct RIXS at the $L$-edge of Ir$^{4+}$ ions in an octahedral crystal field. In the physical limit of strong spin-orbit coupling, the RIXS spectral weight at the $L_2$ vanishes, but it is strong at the $L_3$-edge. Applying this to Sr$_2$IrO$_4$, we find that RIXS can map out the strongly dispersive single- and
double-magnon excitations of the low-lying doublet and is in addition sensitive to the doublet-quartet excitations at an energy of $\tfrac{3}{2}\lambda$, which strongly mix with delocalized charge modes. This shows that RIXS can accurately determine the material parameters $\lambda$ and $J$ of iridates and is an excellent tool to probe their low-energy elementary excitations, testing and characterizing proposed long-range order and topological phases.

We thank H. Takagi, J.~H. Kim, B.~J. Kim and M. van Veenendaal for valuable discussions.


\begin{thebibliography}{99}
\bibitem{AxesFootnote} We refer to the cubic axes, parallel to neighboring Ir-Ir bonds. The $z$ direction is perpendicular to the IrO$_2$ plane.
\end{thebibliography}
\end{document}